\documentclass[
longbibliography,
showpacs,
floatfix,
aps, 
prl,
twocolumn,
superscriptaddress,
amssymb,
tightenlines
]{revtex4-2}
\usepackage{amsmath}
\usepackage{tabulary}
\usepackage{tabularx}
\usepackage{url}
\usepackage[breaklinks=true]{hyperref}
\usepackage{graphicx}
\usepackage{times}
\usepackage{wasysym}
\usepackage{amssymb}
\usepackage{latexsym}

\usepackage{dcolumn}
\usepackage{amsfonts}
\usepackage{bm}
\usepackage{epsfig}

\newcommand{\be}{\begin{equation}}
\newcommand{\ee}{\end{equation}}
\newcommand{\bea}{\begin{eqnarray}}
\newcommand{\eea}{\end{eqnarray}}

\newcommand{\pc}{{\mathcal P}}

\def\nn{\nonumber}
\def\betao{\beta_\omega}
\def\Edc{\mathcal{E}_{\rm dc}}
\def\Eac{\mathcal{E}}

\def\m{m_\star}
\def\eac{\epsilon}
\def\edc{\epsilon_{\mathrm{dc}}}

\def\epseff{\varepsilon_{\mathrm{eff}}}

\def\oc{\omega_{\mbox{\scriptsize {c}}}}

\def\ttr{\tau}
\def\tem{\tau_{\rm em}}
\def\tsh{\tau_{\rm sh}}

\newcommand{\req}[1]{Eq.\,(\ref{#1})}

\newcommand{\rfig}[1]{Fig.\,\ref{#1}}

\newcommand{\rref}[1]{Ref.\,\onlinecite{#1}}
\newcommand{\rrefs}[2]{Refs.\,\onlinecite{#1},\,\onlinecite{#2}}

\def\ne{n_e}

\def\f{f}
\def\g{g}

\def\tcr{\tau_{\rm cr}}

\begin{document}
\title{Oscillatory photoresistance on the high field side of the cyclotron resonance}
\author{M. A. Zudov}
\email[Corresponding author: ]{zudov001@umn.edu}
\affiliation{School of Physics and Astronomy, University of Minnesota, Minneapolis, Minnesota 55455, USA
}
\received{\today}

\begin{abstract}
We consider the displacement contribution to photoresistance in overlapping Landau levels at radiation frequencies much smaller than the cyclotron frequency. 
We show that in the limit of short-range disorder and high radiation power, this contribution leads to another class of magneto-resistance oscillations.
These oscillations, which we call radiowave-induced resistance oscillations (RIROs), are distinct from the well known microwave-induced resistance oscillations in the following aspects: (i) their amplitude is independent of power, (ii) their period is controlled by the radiation electric field, rather than by the radiation frequency, and (iii) they can be either $1/B$ or $1/B^2$-periodic, depending on $B$, with the crossover point linked to the width of the cyclotron resonance absorption curve.
We also show that RIROs should be readily observed in experiments.
\end{abstract}
\maketitle

The field of quantum transport in very high Landau levels was initiated by the discoveries of microwave- \cite{zudov:2001a,ye:2001}, phonon- \cite{zudov:2001b,hatke:2011d}, and Hall (dc) field- \cite{yang:2002,zhang:2007a,zhang:2007b} induced resistance oscillations, followed by observations of zero-resistance \cite{mani:2002,zudov:2003}, zero-conductance \cite{yang:2003}, and zero-differential resistance \cite{bykov:2007,zhang:2008,hatke:2010a} states. 
Among these phenomena, microwave-induced resistance oscillations (MIROs) have received most attention and many theoretical proposals have been put forward \cite{durst:2003,lei:2003,dmitriev:2003,shi:2003, vavilov:2004,dmitriev:2005,dmitriev:2007,dmitriev:2009a,dmitriev:2009b,alekseev:2021}.
Mainstream theoretical models are based on, so-called, ``displacement'' \cite{durst:2003,lei:2003,vavilov:2004} and ``inelastic''\citep{dmitriev:2005,dorozhkin:2003} mechanisms.
To our knowledge, the most general theory for microwave photoresistance was presented in \rref{hatke:2011e}, as it is applicable for arbitrary radiation power.
Here, we use their framework to examine a regime of $\omega \ll \oc$, a high-field side of the cyclotron resonance, whereas MIROs occur at $\omega \gtrsim \oc$.

The photoresistivity $\delta \rho_\omega$ due to displacement contribution \citep{note:sm} in the limit of short-range (``sharp'') disorder in overlapping Landau levels is given by \cite{hatke:2011e,shi:2017a}
\be
\label{eq.pr}
\frac{\delta\rho_\omega}{\rho_D}= 2 \lambda^2 \frac{\tau}{\tsh}  \left[ f(\xi)+\pi\eac\cot\pi\eac\;\xi f'(\xi)\right]\,,
\ee
(need to check the factor of 2) where $\rho_{D}$ is the Drude resistivity, $\tau$ is the momentum relaxation time, $\tsh^{-1}$ is the sharp disorder scattering rate, $\lambda = \exp(-\pi/\mu_q B) \lesssim 1$ is the Dingle factor, $\mu_q$ is the quantum mobility, $B$ is the magnetic field, $\eac = \omega/\oc$, $\omega = 2 \pi f$ is the radiation frequency, $\oc = eB/\m$ is the cyclotron frequency, $\m$ is the effective mass, 
\be \label{eq.gb}
\f(\xi) = J_0^2(\xi) - J_1^2(\xi)\,,
\nn
\ee
where $J_0, J_1$ are Bessel functions of the first kind, and
\be
\xi (\eac) = 2 \sqrt{\pc (\eac)} \sin \pi \eac \,.
\label{eq.xi}
\ee
For linearly polarized radiation the dimensionless radiation power $\pc(\eac)$ in the above equation is given by \citep{khodas:2008}
\be
\pc(\eac) = \frac {\pc_0} 2 \sum\limits_\pm\frac{1}{(1 \pm \eac^{-1})^2+\betao^2}\,,
~\pc_0=\frac{2\pi \ne e^2\Eac^2}{\epseff \m^2 \omega^4}\,,
\label{eq.pc}
\ee
where $\betao = (\omega\tcr)^{-1} =  (\omega\tem)^{-1}+(\omega\ttr)^{-1}$ determines the width of the cyclotron resonance, $\tem^{-1}=(\ne e^2/2\m) \sqrt{\mu_0/\epseff \epsilon_0}$ \citep{chiu:1976,zhang:2014} is the radiative decay rate, $\ne$ is the carrier density, $\sqrt{\epseff}=(\sqrt{\varepsilon}+1)/2$, $\varepsilon$ is the dielectric constant of the surrounding media, and $\Eac$ is the radiation electric field.

Before discussing the new regime, $\eac \ll 1$, we illustrate how \req{eq.pr} explains already observed phenomena.
At $\xi \ll 1$ and $\pi \eac \gg 1$, the oscillatory phootoresistance from \req{eq.pr} becomes \cite{dmitriev:2012}
\be
\frac{ \delta \rho_\omega }{ \rho_{D} } \approx - \frac{3\tau}{\tsh}  \pc(\eac) \lambda^{2} 2 \pi \eac  \sin 2\pi\eac\,,
\label{eq.miro}
\ee
which accounts for MIROs \cite{zudov:2001a,ye:2001}.
According to \req{eq.miro}, MIROs are periodic in $B^{-1}$, with the amplitude proportional to $\pc$ and the frequency controlled by $\omega$.
Similarly, at $\xi \gg 1$ and $\pi \eac \gg 1$, \req{eq.pr} can be approximated as
\be
\frac{ \delta \rho_\omega }{ \rho_{D} } \approx \frac{8 \tau}{\tsh}\lambda^2 \eac \cot (\pi\eac) \cos \left [4 \sqrt{\pc(\eac)}\sin \pi\eac \right ]\,,
\label{eq.fs}
\ee
accounting for the fine structure of MIROs \cite{shi:2017a}.
Experimental observation of the fine structure \cite{shi:2017a} confirmed the general validity of the theory and offered a method to probe microwave electric field entering $\pc$.
However, the residual dependencies of both the amplitude and the period on $\eac$ considerably complicates the analysis \cite{shi:2017a}.

Here, we focus on the regime of $\xi \gg 1$ and $\pi \eac \ll 1$.
When $\xi \gg 1$, $\f(\xi)$ and its derivative $\f'(\xi)$ can be expressed as 
\be \label{eq.gba}
\f(\xi) \approx \frac {2\sin 2\xi} {\pi \xi}\,,~~
\f'(\xi) \approx \frac {4\cos 2\xi} {\pi \xi}\,.
\nn
\ee 
At $\pi \eac \ll 1$, we can recast $\pc$ and $\xi$ as
\be
\pc(\eac) \approx {\pc_0\eac^2} \g^2(\betao\eac)\,,~~ 
\xi (\eac) \approx 2 \pi \sqrt{\pc_0} \eac^2  \g(\betao\eac) \,,
\label{eq.pc.a}
\nn
\ee
where $\g(x) = \left(1 +x^2\right )^{-1/2}$.
We thus see that the first term in \req{eq.pr} can be ignored and, since $\pi \eac \cot \pi\eac \approx 1$,
the photoresistivity can be written as
\be
\frac{ \delta \rho_\omega }{ \rho_{D} } \approx \frac{8 \tau}{\pi\tsh} \lambda^2 \cos 2\pi \kappa\,,
\label{eq.result}
\ee
where
\be
\kappa = 
\frac{ \Eac_\star \m} { e B^2}  \frac{2k_F}{\sqrt{1 + (\oc\tcr)^{-2}}}\,,
\label{eq.kappa}
\ee
$\Eac_\star = \Eac/\sqrt{\epseff}$, $(\oc\tcr)^{-1} = \betao\eac$, and $k_F = \sqrt{2\pi\ne}$.

We will refer to the oscillations described by \req{eq.result}, which is our main result, as radiowave-induced resistance oscillations (RIROs) since, as we show below, they are likely to be best observed at radiation frequencies of $f \lesssim 1$ GHz.
We note that the properties of RIROs are independent of $\omega$, and their frequency is proportional to the radiation field $\Eac$.

We next show that RIROs should be readily detected in experiment.
For this purpose, we consider a two-dimensional (2D) system based on GaAs/AlGaAs quantum well with electron density $\ne = 3 \times 10^{15}$ m$^{-2}$ and mobility $\mu = 3 \times 10^3$ m$^2$/Vs. 
We further use $\tau/\tsh = 0.5$ \citep{sammon:2018}, the quantum mobility $\mu_q = 1 \times 10^2$ m$^2$/Vs \citep{shi:2017a}, the dielectric constant of GaAs $\varepsilon = 12.8$, and the effective mass $\m  = 0.06$ $m_0$ \citep{hatke:2013}, where $m_0$ is the mass of free electron. 
We also fix the radiation electric field, which determines real radiation power, at a conservative value of $\Eac = 250$ V/m (unless otherwise noticed). 

\begin{figure}[t]
\includegraphics{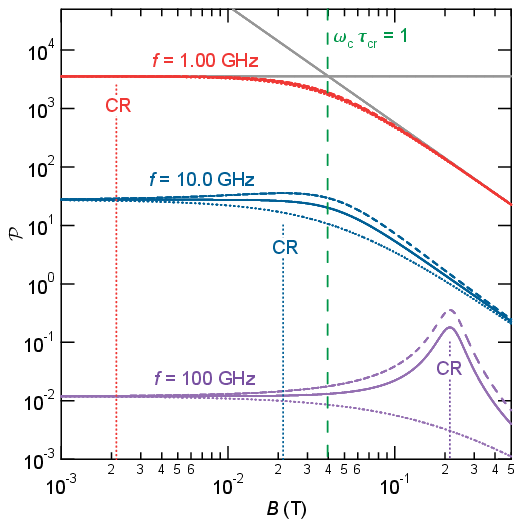}
\vspace{-0.15 in}
\caption{\small{
Dimensionless microwave power $\pc$ as a function of magnetic field $B$, computed using \req{eq.pc}, for $f = 1.00$, 10.0, and 100 GHz, as marked.
Solid lines correspond to linear polarization, \req{eq.pc}, and dashed (dotted) lines to circular active (inactive) polarizations.
Dotted vertical lines mark the position of the cyclotron resonance (CR) $\oc = \omega$.
Dashed vertical line is drawn at $B = m^\star/e\tcr$.
}}
\label{fig1}
\vspace{-0.2 in}
\end{figure}

At $\eac \ll 1$, the effective radiation power $\pc$ scales with $\omega^{-2}$, per \req{eq.pc}, which suggests that lowering the radiation frequency helps to satisfy the condition $\xi \gg 1$.
In \rfig{fig1} we show $\pc$ as a function of the magnetic field $B$ for three frequencies, $f = 100$, 10.0, and 1.00 GHz, as marked.
Here, the solid lines correspond to linear polarization, \req{eq.pc}, and the dashed (dotted) lines are the results for circular active (inactive) polarizations.
The vertical lines are drawn at $\oc = \omega$ (dotted) and at $\oc = \tcr^{-1}$, as marked.
At the lowest  frequency, $f = 1.00$  GHz, it is clearly seen that (i) $\pc \gg 1$ over a wide range of $B$ corresponding to $\eac \ll 1$, (ii) the crossover from $\pc \propto B^0$ to $\pc \propto B^{-1}$ takes place at $\oc  = \tcr^{-1} \gg \omega$, and (iii) the sense of the radiation polarization becomes irrelevant. 
For the intermediate frequency, $f = 10.0$ Ghz, we observe that $\pc \gg 1$ only when $\eac \gtrsim 1$.
At the highest frequency, however, we find that $\pc \ll 1$ over the whole range of $B$.

\begin{figure}[t] 
\includegraphics{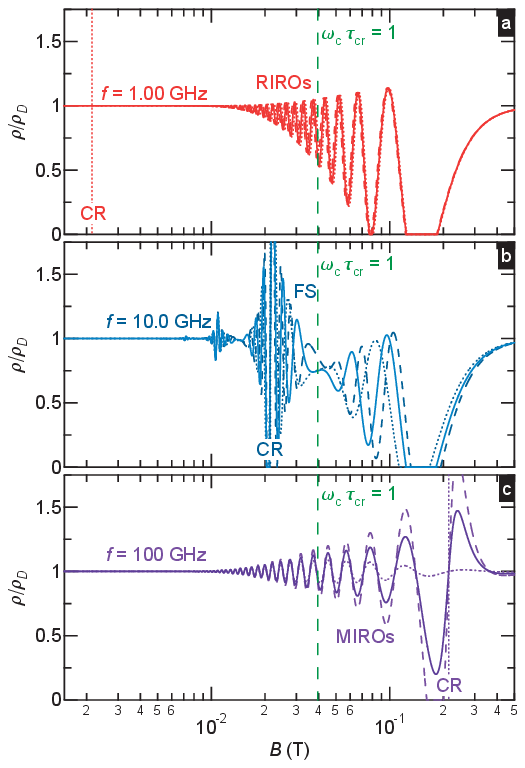}
\vspace{-0.15 in}
\caption{\small{
Photoresistance $\rho/\rho_D$ as a function of magnetic field $B$ calculated using \req{eq.pr} for (a) $f = 1.00$ GHz, (b) 10.0 GHz, and (c) 100 GHz, as marked \citep{note:1}.
Solid lines correspond to linear polarization, \req{eq.pc}, and dashed (dotted) are the results for circular active (inactive) polarizations.
Dotted vertical lines mark the position of the cyclotron resonance (CR) $\oc = \omega$.
Dashed vertical line is drawn at $B = m^\star/e\tcr$.
}}
\label{fig2}
\vspace{-0.2 in}
\end{figure}
We next compute the photoresistance using \req{eq.pr} for the same set of frequencies, $f = 1.00$, 10.0, and 100 GHz, and present the results in \rfig{fig2}.
At the highest frequency, $f = 100$ GHz (panel c), the photoresistance exhibits MIROs \cite{zudov:2001a}, which occur on the lower field side of the cyclotron resonance and are well described by \req{eq.miro}.
At the intermediate frequency, $f = 10.0$ GHz (panel b), we see more complex behavior, which includes the fine structure of MIROs (marked FS), observed near $\eac = 1$ (and $\eac = 2$).
This fine structure occurs when $\xi \gg 1$ and can be described by \req{eq.fs} \cite{shi:2017a}.

At the lowest frequency, $f = 1.00$ GHz (panel a), the photoresistance oscillation pattern simplifies considerably, resembling that of MIROs. 
However, even though these RIROs occur in the same range of magnetic fields as MIROs (panel c), much lower $f$ ensures that $\pi \eac \ll 1$.
As discussed above, RIROs are independent of $\omega$, which we have confirmed by calculations at $f = 0.10$ GHz, see \rfig{fig4}(a).
This is indeed in agreement with our main result, \req{eq.result} and \req{eq.kappa}.

\begin{figure}[t] 
\includegraphics{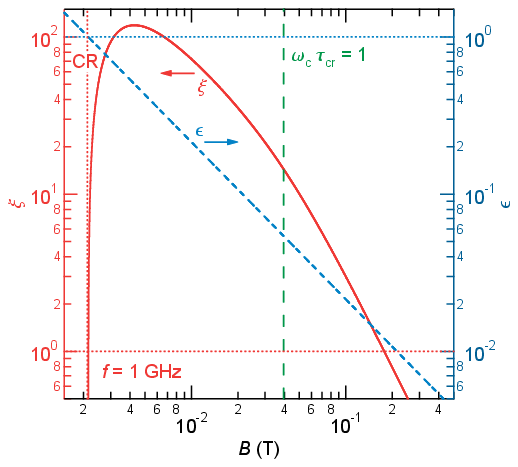}
\vspace{-0.15 in}
\caption{\small{
Parameter $\xi$, \req{eq.xi}, (solid line, left axis) and $\eac = \omega/\oc$ (dashed line, right axis) as a function of magnetic field $B$ calculated for $f = 1.00$ GHz.
Dashed vertical line is drawn at $B = m^\star/e\tcr$.
}}
\label{fig3}
\vspace{-0.2 in}
\end{figure}

\begin{figure}[t] 
\includegraphics{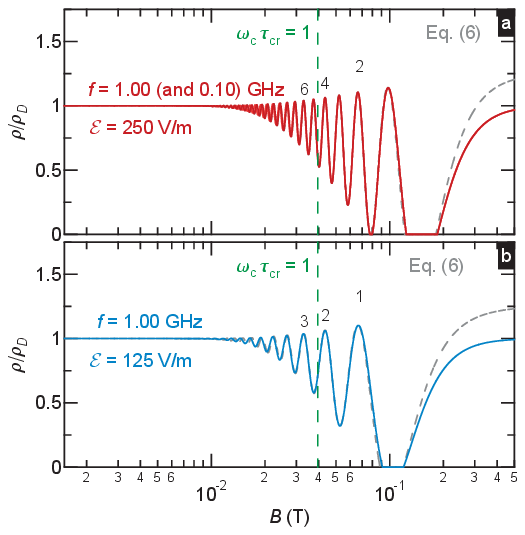}
\vspace{-0.15 in}
\caption{\small{
(a) Photoresistivity $\rho/\rho_D$ as a function of magnetic field $B$ calculated using \req{eq.pr} (solid lines) and \req{eq.result} (dashed lines) for $f = 1.00$ GHz (and for $f = 0.10$ GHz) \citep{note:1}.
The microwave field is $\Eac = 250$ V/m.
As expected, the curves corresponding to different $f$ are indistinguishable.
(b) Photoresistivity $\rho/\rho_D$ as a function of magnetic field $B$ calculated using \req{eq.pr} (solid line) and \req{eq.result} (dashed line) for $f = 1.00$ GHz and $\Eac = 125$ V/m \citep{note:1}.
The oscillation frequency is reduced by a factor of two, compared to RIROs shown in panel (a), but the amplitude remains the same.
Approximations given by \req{eq.result} break down at high $B$,  when $\xi$ is no longer much larger than unity, see \rfig{fig3}.
Dashed vertical line is drawn at $B = m^\star/e\tcr$.
}}
\label{fig4}
\vspace{-0.2 in}
\end{figure}

In the derivation we have assumed that $\xi \gg 1$, which is a more stringent condition than $\pc \gg 1$, when $\pi\eac \ll 1$, see \req{eq.xi}.
As shown in \rfig{fig3}, for $f = 1.00$ GHz, both $\xi \gg 1$ (left axis) and $\eac \ll 1$ (right axis) conditions are satisfied for magnetic fields between $B \simeq 0.01$ T and $B \simeq 0.1$ T, a range where RIROs are seen in \rfig{fig2}(a).
As shown in \rfig{fig4}, our approximation, \req{eq.result} (dashed line) is indeed very good up to $B \simeq 0.1$ T, but breaks down at higher $B$, as expected, because of decreasing $\xi$.
In addition, comparison of \rfig{fig4}(a) and (b) shows that when the radiation field is reduced by a factor of two, the oscillation frequency is reduced by a factor of two as well, while the amplitude is not affected.
Finally, we note that the model assumes overlapping Landau levels, $\lambda \ll 1$, so the shape of the oscillations might become distorted at higher $B$ \citep{hatke:2011f,hatke:2012d}.

According to \req{eq.result}, $\oc\tcr$ affects both the amplitude and the frequency of RIROs.
At high magnetic fields, $\oc\tcr \gg 1$, \req{eq.kappa} can be approximated as
\be
\kappa \approx \frac{\Eac_\star \m (2k_F)} {e B^2}\,,
\label{eq.result.h}
\ee
which results in $B^{-2}$-periodic oscillations.
In the opposite limit,  $\oc\tcr \ll 1$, one finds 
\be
\kappa \approx \frac{\Eac_\star\tcr(2k_F)}{B}\,,
\label{eq.result.l}
\ee
and the oscillations are periodic in $B^{-1}$.
The crossover between these two regimes takes place at $\oc\tcr = 1$ ($B = m^\star/e\tcr$).
As illustrated in \rfig{fig2}, the crossover point lies in the middle of the magnetic field range where RIROs develop and therefore should also be detectable in experiments.  

It is established that experiments at high radiation intensities are limited to elevated temperatures, $T \gtrsim 1.5$ K \citep{khodas:2010,hatke:2011e,shi:2017a}, due to heating by the radiation. 
It is thus anticipated that RIROs will decay with increasing power, as a result of enhanced electron-electron scattering rate entering the quantum scatting rate in the Dingle factor \citep{hatke:2009a,shi:2017a}.
Nevertheless, experimental observation of the fine structure \citep{shi:2017a} strongly suggests that RIROs should also be observed. 
We also note that one should not anticipate any interference with  Shubnikov-de Haas oscillations \citep{shi:2015b} at such  temperatures.

It is also understood that the inelastic contribution is suppressed both by high temperature and by high radiation power \citep{hatke:2011e,shi:2017a}, while the smooth-disorder displacement contribution does not give rise to oscillatory photoresistance at $\eac \ll 1$. 
These contributions are discussed in Supplemental Material.

We also would like to mention that RIROs resemble dc field-induced resistance oscillations \citep{yang:2002,zhang:2007a,bykov:2012}.
These oscillations are due to elastic electron transitions between Landau levels, tilted by dc electric field $\Edc$, due to impurity scattering.
The rate for such transitions is maximized each time when a spatial separation between Landau levels matches the cyclotron diameter $2R_c$, a condition for electron back scattering.
The dc field-induced-oscillations are described by \citep{vavilov:2007}
\be
\frac{ \delta r }{ \rho_{D} } \approx \frac {16\tau}{\pi\tsh} \lambda^2 \cos 2\pi \epsilon_{\rm dc}\,,~~\epsilon_{\rm dc} =  \frac{e \Edc (2R_c)}{\hbar \oc}\,,
\label{eq.hiro}
\ee
where $\delta r$ is the correction to the differential resistivity.
Since $R_c = \hbar k_F/e B$, $\oc = eB/\m$, the parameter $\kappa$ in \req{eq.result.h} can also be written in the same form as $\edc$, $\kappa = e \Eac_\star (2R_c)/\hbar \oc$. 
We thus see that \req{eq.result} and \req{eq.hiro} have the same form, up to numerical factor, which suggests that these two quite distinct phenomena share the same underlying physical mechanism.
However, as discussed in \rrefs{hatke:2011e}{shi:2017a}, RIROs can also be naturally understood in terms of multi-photon absorption.

In summary, we have shown that the displacement contribution in the sharp disorder limit gives rise to a distinct class of magneto-oscillations, RIROs, which appear on the high-field side of the cyclotron resonance under irradiation by radiowaves.
All properties of RIROs are independent of the radiation frequency, reflecting the fact that the radiowave electric field $\Eac$ appears essentially static for electrons when $\omega \ll \oc$.
While the amplitude of RIROs does not depend on the intensity or polarization of the radiowaves, their frequency scales with the radiation electric field.
Finally, we identified two regimes in which RIROs are periodic in either $1/B$ or $1/B^2$, with the crossover point governed by the width of the cyclotron resonance.
Unique characteristics of RIROs should allow mapping the cyclotron resonance absorption curve and accurate determination of $\Eac$.
As MIROs \cite{konstantinov:2009,zudov:2014,karcher:2016,otteneder:2018,monch:2020,savchenko:2020}, this new phenomenon can likely be realized in a variety of 2D systems.

\begin{acknowledgments}
The author thanks D. Polyakov for discussions and acknowledges support by the NSF Grant No. DMR-1309578.
\end{acknowledgments}


\end{document}